# Highly efficient visible colloidal lead-halide perovskite nanocrystal light-emitting diodes

Fei Yan[1,‡], Jun Xing[2,3,‡], Guichuan Xing[4], Lina Quan[3], Swee Tiam Tan[1], Jiaxin Zhao[2], Rui Su[2], Lulu Zhang[2], Shi Chen[2], Yawen Zhao[5], Alfred Huan[2], Edward H. Sargent[3*], Qihua Xiong[2,6*], and Hilmi Volkan Demir[1,2,7*]

[1] LUMINOUS! Center of Excellence for Semiconductor Lighting and Displays, TPI-The Photonics Institute, School of Electrical and Electronic Engineering, Nanyang Technological University, Singapore 639798,

[2] Division of Physics and Applied Physics, School of Physical and Mathematical Sciences, Nanyang Technological University, Singapore 637371,

[3] Department of Electrical and Computer Engineering, University of Toronto, 10 King's College Road, Toronto, Ontario M5S 3G4, Canada,

[4] Institute of Applied Physics and Materials Engineering, University of Macau, Macao SAR 999078, China,

[5] China Academy of Engineering Physics, Mianyang 621900, China,

[6] NOVITAS, Nanoelectronics Centre of Excellence, School of Electrical and Electronic Engineering, Nanyang Technological University, Singapore 639798,

[7] Department of Electrical and Electronics Engineering, Department of Physics, UNAM−Institute of Materials Science and Nanotechnology, Bilkent University, Ankara 06800, Turkey.

‡These authors contributed equally to this work.

*e-mail: volkan@stanfordalumni.org, qihua@ntu.edu.sg, ted.sargent@utoronto.ca

Lead-halide perovskites have been attracting attention for potential use in solid-state lighting. Following the footsteps of solar cells, the field of perovskite light-emitting diodes (PeLEDs) has been growing rapidly. Their application prospects in lighting, however, remain still uncertain due to a variety of shortcomings in device performance including their limited levels of luminous efficiency achievable thus far. Here we show high-efficiency PeLEDs based on colloidal perovskite nanocrystals (PeNCs) synthesized at room temperature possessing dominant first-order excitonic radiation (enabling a photoluminescence quantum yield of 71% in solid film), unlike in the case of bulk perovskites with slow electron-hole bimolecular radiative recombination (a second-order process). In these PeLEDs, by reaching charge balance in the recombination zone, we find that the Auger nonradiative recombination, with its significant role in emission quenching, is effectively suppressed in low driving current density range. In consequence, these devices reach a record high maximum external quantum efficiency of 12.9% reported to date and an unprecedentedly high power efficiency of 30.3 lm $W^{-1}$ at luminance levels above 1000 cd $m^{-2}$ as required for various applications. These findings suggest that, with feasible levels of device performance, the PeNCs hold great promise for their use in LED lighting and displays.

Lead-halide perovskites with direct band gaps have emerged as a new material system for low-cost, large-area and light-weight optoelectronics[1-21]. Previous reports have confirmed promising performance for this material family in light generation[13-21]. With binding energy up to dozens of meV and long electron-hole diffusion length of micrometers at room temperature, carriers are almost free and delocalized in bulk perovskites[16,22-24] leading to slow electron-hole bimolecular radiative recombination (a second-order process)[11,24]. Competing with trap-mediated nonradiative decay (a first-order process), this radiation process limits the efficiency levels[11,24]. Recently a strategy that enhances the photoluminescence quantum yield (PLQY) of light emitters and external quantum efficiency (EQE) of their LEDs has been developed by localizing the charge carriers in small sized lead-halide perovskite grains or quantum wells[16-21]. However, even though such colloidal perovskite nanoparticles can present PLQY of near unity in solution and above 60% in film, their performance levels have been still severely limited during the device operation to date[19,20,25,26]. Therefore, deeper understanding of device performance limits and a new way divorced from this low performance are required. Here we present highly efficient PeLEDs based on colloidal $CH_3NH_3PbBr_3$ ($MAPbBr_3$) PeNCs with efficient first-order excitonic radiation reaching a 71% PLQY in film. Due to the significant role of Auger nonradiative recombination, highly efficient emission at low charge densities is desired. By effectively suppressing Auger nonradiative recombination with balanced charge at low driving charge density levels, these PeLEDs exhibit a record high EQE of 12.9% up to date and a high power efficiency level of 30.3 lm $W^{-1}$ and high brightness level above 1000 cd $m^{-2}$, which indicates the feasibility of colloidal PeNCs in LED applications in terms of efficiency requirements.

**Luminescence dynamics of lead-halide perovskite emitters**

The radiation of bulk perovskites originates from slow bimolecular recombination. The Langevin theory predicts that this bimolecular recombination rate could be four orders-of-magnitude higher in perovskites, which is one of the main merits ensuring superior performance in photovoltaics[11,24,27-30]. The carrier kinetics in bulk lead-halide perovskites can be described by:

$$\frac{dn(t)}{dt} = G - nk_1 - n^2k_2 - n^3k_3 \tag{1}$$

where $n$ is the carrier density, $t$ is the time, $G$ is the generation rate of carrier density, $k_1$ is the first-order trap-mediated monomolecular recombination rate constant, $k_2$ is the free carrier bimolecular recombination rate constant, and $k_3$ is the three-body Auger recombination rate constant[11,24,27-30]. Thus, under steady-state electric excitation, the luminous QY ($A(n)$) is given by:

$$A(n) = \frac{nk_2}{k_1 + nk_2 + n^2k_3} \tag{2}$$

The emission efficiency (Fig. 1a) first increases with increasing $n$ due to the domination of bimolecular recombination over the trap-mediated nonradiative recombination. However, at high carrier densities, where the Auger recombination dominates over the bimolecular recombination, the emission efficiency will decrease with increasing $n$. Since $k_2$ (~$10^{-10}$ cm$^3$ s$^{-1}$) and $k_3$ (~$10^{-28}$ cm$^6$ s$^{-1}$) are the intrinsic parameters of the bulk perovskites[11,24,27-30], the emission efficiency strongly depends on $k_1$ and $n$. Ignoring other quenching factors, with a typical reported value of $k_1$ (~$10^7$ s$^{-1}$, 100 ns), the internal efficiency of the PeLEDs is constrained to <10%, even when driving at a high carrier density of $10^{16}$ cm$^{-3}$ (Fig. 1a).

The bulk perovskite films have shown a long carrier lifetime (i.e., low trap states density)[31] and a high mobility[32]. Thus, the delocalized charge carriers are also possible to transport through the thin active layer without recombination. When the PeLEDs are driven at a lower carrier density level or under a higher trap density regime (trap-related recombination time <100 ns), the nonradiative recombination could always dominate over the bimolecular recombination, which also leads to an internal QY of <10%. Therefore, the slow bimolecular recombination rate in bulk perovskites limit the PeLEDs working at a relatively low carrier density. The quadratic power dependence of the recombination rate on the carrier densities of MAPbBr$_3$ bulk crystal (Fig.1b) confirms the bimolecular recombination in the bulk perovskite. The experimentally measured curves match well with the theoretical prediction at the low charge density level (Fig. 1a, c). It is noted that the first-order excitonic emission could compete effectively with the trap-mediated nonradiative recombination in the low carrier density range of $n \leq 10^{16}$ cm$^{-3}$ with trap states density ($k_1 \sim 10^7$ s$^{-1}$) as shown in Fig.1a. Therefore, small-size PeNCs can enhance the radiative excitonic recombination with increased exciton confinement and exciton binding energy comparing to bulk crystals with dominating free-carrier recombination. Moreover, the surrounded passivation of ligands can decrease the surface defects of PeNCs, which also leads to an increment of QY at low excitation density level.

**Synthesis of PeNCs and their luminescence dynamics**

The colloidal $MAPbBr_3$ PeNCs were synthesized by using anti-solvent precipitation method at room temperature. The as-synthesized solution was pre-aged for 6 days before being used for LEDs fabrication. In this duration, the absorption edge and PL peak gradually red-shift until after 6 days, and the full-width-at-half-maximum (FWHM) of PL spectra keeps narrowing synchronously (Fig. 1d, Supplementary Fig. 1 ). Notably, the PLQY was significantly enhanced from 70% in solution and 35% in film to 89% in solution and 71% in film. From nuclear magnetic resonance (NMR), the composition of the $MAPbBr_3$ NCs was found to remain stable during the aging time (Supplementary Table 1). However, as observed in X-ray diffraction (XRD) patterns, the $MAPbBr_3$ NCs show increased crystallization during this process (Fig. 1e). The high-resolution transmission electron microscopy (HR-TEM) measurements (Fig. 1f) illustrate that the aged PeNCs still exhibit a small crystal size (< 8 nm). The lattice fringes of the $MAPbBr_3$ PeNCs are 2.93 and 2.65 Å, which correspond to the (200) and (210) facets of the cubic phase $MAPbBr_3$, respectively. With the passivation of ligands and solvent molecules, the low-crystallinity $MAPbBr_3$ NCs in colloidal solution possess a high PLQY. However, the passivation effect decreases drastically in film (Supplementary Table 1), thus the defect sites become significant in quenching emission through the competition with radiation process. Though with similar passivation, the highly crystallized NCs with a low defect sites display low trap-assisted nonradiative recombination, which ensures a high PLQY in film. Even with such passivation effects, the PeNCs still shows high mobility of 0.22 $cm^2$ $V^{-1}$ $S^{-1}$ for holes and 0.48 $cm^2$ $V^{-1}$ $S^{-1}$ for electrons (Supplementary Fig. 2).

As illustrated in Fig. 1b, for MAPbBr$_3$ PeNCs, I$_{PL}$[t = 0] is nearly linearly dependent on the carrier density from 5 × 10$^{14}$ to 2 × 10$^{17}$ cm$^{-3}$, which suggests that the emission is mainly from the first-order excitonic radiation process. Below 5 × 10$^{14}$ cm$^{-3}$, the trap-mediated nonradiative recombination dominates over the excitonic emission. Above 2 × 10$^{17}$ cm$^{-3}$, the multi-particle Auger recombination prevails. Therefore, MAPbBr$_3$ PeNCs feature a high PLQY level under the charge carrier density from 10$^{15}$ to 10$^{16}$ cm$^{-3}$, which matches well with the device working regime in PeLEDs (Fig. 1c)[11,12]. Therefore, the highly efficient first-order of excitonic radiation enables the MAPbBr$_3$ PeNCs as a suitable candidate possibly for efficient PeLEDs.

**Investigation of PeLEDs**

As the topmost important parameter for EL device characterization, EQE, which is defined as the number of emitted photons per injected electrons, is given as the product of four different parameters:

$$EQE = E_{in} \cdot E_{s/t} \cdot E_{rad} \cdot E_{out} \qquad (3)$$

Therein the first parameter $E_{in}$ indicates the charge balance factor in recombination. The second one $E_{s/t}$ gives the fraction of excitons that are allowed to radiate by spin statistics. The third factor $E_{rad}$ depicts the fraction of the intrinsic radiative efficiency of the emitter. The last one $E_{out}$ determines the photon extraction efficiency in a device. In this work with the proposed devices using normal structure without any additional measurement for spin injection of charge carriers or enhanced outcoupling efficiency of radiation, thus all attention was paid onto the charge balance factor in recombination zone to reach the high device efficiency promised by MAPbBr$_3$ PeNCs. Moreover, due to the significant role of Auger nonradiative recombination (Fig. 1a, c) which is to be effectively suppressed in the proposed MAPbBr$_3$ PeNCs devices, a high EQE in low driving current density level was also targeted. In addition to EQE, the power efficiency, especially at high brightness level, is also a critical parameter for evaluating the potential applications of LEDs; therefore, a low driving voltage is essential.

To obtain high quality PeNCs film, a classical PEDOT:PSS layer was selected as hole injection/transport layer (HIL/HTL) for all devices in this work (Fig. 2a, b, Supplementary Fig. 3, 4).

As a good HIL/HTL, PEODT:PSS may lead to emission quenching of PeNCs at HTL/PeNCs interface, but we did not observe a significant weakness of emission comparing to the PeNCs film deposited onto other polymer substrates, like poly-TPD, at an excitation wavelength of 365 nm. The insertion of an insulating thin-film can improve charge balance by blocking the majority carriers, but this would probably be accompanied with high driving voltage, which would lead to high power consumption and low power efficiency.

To balance the injected hole, a blended layer of B3PYMPM:TPBi (1:X w/w) with adjustable electron transport capability by changing the ratio of two components served as an electron transport layer (ETL) (Fig. 2a, b). The deeper LUMO level (-3.4 eV) and high electron conductivity ($4.1 \times 10^{-7}$ S cm$^{-1}$) of B3PYMPM can lower down the barrier for effective electron injection and transport with low driving voltage[20,33,34]. Due to the significant difference of the electron conductivity in TPBi and B3PYMPM[33], the electron transport capability of the blended ETL can be tuned in a large range (Fig. 2c). According to Miller–Abrahams hopping model, electrons transport through a hopping process between the localized occupied and unoccupied state. In the blended ETL, TPBi (B3PYMPM) molecules play the role of an energy barrier (trap) to scatter the electron transfer between two B3PYMPM (TPBi) molecules, leading to a tunable electron transport capability[35]. In the single carrier device with TPBi ETL, as the low electron conductivity ($4.2 \times 10^{-10}$ S cm$^{-1}$) and high energy barrier for electron injection caused by the high lying LUMO level of TPBi, the driving voltage is much higher than that of single carrier device with B3PYMPM ETL, which can offer a much higher level of electron conductivity and a deeper LUMO level (Fig. 2a, c)[33,34]. For the single carrier device with B3PYMPM:TPBi (1:X, w/w) ETL, the blended layer shows a tunable electron transport capability depending on the varied component ratio (Fig. 2c). To reach a high power efficiency a n-type doped electron injection layer was used for a low driving voltage[20].

Control device I with pure TPBi and control device II with pure B3PYMPM were selected as the baseline. As seen in Fig. 2d, PeLEDs show a green color emission with a nearly ideal Lambertian profile (Supplementary Fig. 5). Control device I exhibits overall performance of 15,130 cd m$^{-2}$ in $L_{max}$, 5.11 % in EQE$_{max}$, and 10.82 lm W$^{-1}$ in η$_{max}$ (Figs. 2, 3, Table 1, and supplementary Table 2). On the other hand, control device II displays a higher L$_{max}$ of 24,410 cd m$^{-2}$ and a drastically decreased driving voltage, but a lower η$_{max}$ of 5.96 lm W$^{-1}$ caused by a lower EQE$_{max}$ of 2.39% compared to control device I (Table 1 and Figs. 2, 3). With B3PYMPM:TPBi (1:2, w/w) ETL control device III shows the best overall performance up to now, with L$_{max}$ of 33,570 cd m$^{-2}$, EQE$_{max}$ of 7.91%, and η$_{max}$ of 20.18 lm W$^{-1}$ (Figs. 2, 3, Table 1, and supplementary Table 3). In all devices, the influence of $E_{s/t}$, $E_{rad}$, and $E_{out}$ can be neglected because the emitters are identical and the difference in refractive index between TPBi and B3PYMPM is also negligible. Therefore, the significant difference in EQE$_{max}$ among control devices is primarily caused by different charge balance in the recombination zone.

Generally, the brightness in a LED is correlated with the amount of activated emitters in the recombination zone. In all control devices, the nanocrystals distribution should be the same, hence the higher brightness means more activated emitters in the recombination zone. As the driving current density increases, more emitters were excited, leading to increment of brightness. With the highest charge supply capability of B3PYMPM, control device II, instead, gave the lowest luminous efficiency among all devices, even though its L$_{max}$ is higher than that of control device I by 61%. As the emitters were the same for all devices, the possible reason is the imbalanced charge in the recombination zone of control device II. Additionally, the red emission in Ir(piq)$_2$acac indicator device with TPBi ETL suggested the injection of remaining hole into ETL as the maximum effective radius of Förster energy is 6 nm. It was thus deducted that hole is the major charge in the recombination zone of control device I (Supplementary Fig. 6). Otherwise, the pure PeNCs green color emission in Ir(piq)$_2$acac indicator device with B3PYMPM ETL would have suggested there was less or no hole injection into ETL comparing to TPBi devices (Supplementary Fig. 6). With the same hole injection, the highest EQE$_{max}$ suggested the best charge balance in control device III with medium electron supply ETL comparing to I and II (Fig. 2, 3). Therefore, the lowest EQE$_{max}$ of control device II was ascribed to the overwhelming majority of electrons in the PeNCs emissive layer as the highest electron supply of B3PYMPM.

In amorphous organic semiconductor the LUMO/HOMO level distribution was broad, of which tail extended into band gap as gap states[36]. Moreover, as $\sigma = ne\mu_e$ (where σ is electron conductivity, $n$ is the electron concentration, $e$ is the elementary charge, $\mu_e$ is the electron mobility) indicates, a comparable electron mobility but significantly high electron conductivity suggest high electron concentration. Thus, here it was deducted that the amount of gap sates in B3PYMPM is much higher comparing to TPBi, which can be supported by the smaller distance between the Fermi level and LUMO level of B3PYMPM comparing to TPBi (Supplementary Fig. 7). In consequence, with a deeper LUMO level, there were more electrons injected into B3PYMPM layer even when the applied bias was less than the threshold voltage of 2.2 V comparing to TPBi (Fig. 2e). By the law of charge conservation, in a closed-loop circuit, the electrons injected into B3PYMPM layer will transit to the valence band of PeNCs by a nonradiative process, leading to leakage current (Fig. 2a, e). Moreover, under high driving voltage the excessive electrons injected into the conductive band of PeNCs will transit to the Fermi level of PEDOT:PSS by a nonradiative process as current overflow, which is another way of leakage current for control device II working in high driving current density range. The leakage of carriers which flow through the device without recombining drives down the EQE directly. The imbalanced charge also leads to an enhanced Auger nonradiative recombination. Thus, control device II presented the lowest $EQE_{max}$ among all devices. As exhibited in Fig. 2e, with blended ETL, the leakage current density was suppressed significantly by 3-4 magnitudes. In most practical LEDs such leakage current was ascribed to device inherent structure. Therefore we believe that, in control device III by decreasing the amount of electron injected into the recombination zone with a blended ETL, the leakage current was suppressed and the charge balance in recombination zone was improved simultaneously.

Furthermore, a thicker PeNCs layer containing more emitters but fewer pinholes will be helpful to obtain a broader recombination zone, higher luminance, and lower leakage current density. As the comparable mobility of hole and electron, the thickness change in PeNCs layer will not influence the charge balance in the recombination zone significantly. Therefore, a thicker (80 nm) PeNCs layer was deposited using a mixed THF:CB (Y:1 v/v, Y=16) PeNCs colloidal solution, by means of which the champion device reached a record $EQE_{max}$ of 12.9 % with the lowest leakage current (Fig. 2e, 3a, Table 1). Assuming the light outcoupling efficiency of around 20%, the estimated internal quantum yield reaches 64.5%[20]. The $EQE_{max}$ histogram for 15 devices from 4 batches describes the distribution of $EQE_{max}$ of the fabricated PeLEDs (Fig. 3d). Due to the significant difference between CB and THF colloidal solution in PeNCs film deposition (Supplementary Fig. 8, 9), the reproducibility of PeNCs film deposition using THF:CB (Y:1 v/v, Y=16) colloidal solution is moderate. To balance the $EQE_{max}$ and $L_{max}$, the peak value of 43,440 cd m$^{-2}$ in $L_{max}$, 12.9% in EQE, and 30.3 lm W$^{-1}$ in η (at a luminance level of 2,069 cd m$^{-2}$, which meets the requirement for lighting application of 1000 cd m$^{-2}$) were observed from different champion devices (Table 1, Fig. 3). Furthermore, we measured the device half-lifetime to be around 6 min (Supplementary Fig. 10), which is still a daunting problem for PeLEDs, and there is a considerable margin to improve. Moreover, with the same device structure except a slightly changed ratio of B3PYMPM:TPBi (1:1.5, w/w), a PeLEDs with emitter of $CsPbBr_3$ nanocrystals also offered a high performances, including a $EQE_{max}$ of 12.4%, which details will be dicussed in the following work.

For all devices it was found that a higher $EQE_{max}$ corresponds to a lower driving current density ($J_c$) (Fig. 3a, e). As the device structures were different, the exponential dependence was probably related to the luminescence dynamics of PeNCs. As the rising of driving current density EQEs increases from very low initial values (Fig. 3a), which is caused by the competition of traps. Additionally, as the high mobilities of PeNCs holes and electrons meet and recombine with low probability, which is also a reason for the negligible EQE value in the low driving current density. With the increasing driving current density, the radiative excitonic recombination takes a dominating role gradually. With the further increase of the driving current density, the Auger nonradiative recombination dominates over the excitonic recombination. However, the imbalanced charges in the recombination zone, which do not exist in the photoexcitation process, boosts the nonradiative Auger process even at the low driving current density level (Fig. 3a, f). As shown by the electron single carrier

PeNCs device, the excessive electrons injected into the PeNCs layer leads to a significantly quenched PL even at a low electron density (n) level. Though the charge balance factors are the same in the recombination zone for control device III and champion device, the thicker emissive layer of champion device can decrease the charge density, allowing for a weak Auger nonradiative recombination compared to control device III.

**Conclusion**

In summary, we demonstrated highly efficient PeLEDs using colloidal organometal halide PeNCs. The high-quality colloidal MAPbBr$_3$ PeNCs with efficient one-order excitonic emission was synthesized at room temperature. As the significant Auger nonradiative recombination caused by high driving current density and imbalanced charge, the PeLEDs with optimized charge balance in low driving current density level were developed and reached a record EQE$_{max}$ of 12.9% up to date accompanied with high $\eta_{max}$ of 30.4 lm W$^{-1}$ at high brightness above 1000 cd m$^{-2}$. This work demonstrates a breakthrough advancement toward the ultimate goal of application of PeLEDs and lays down a systematic constructive approach for researchers on halide PeLEDs in their future works.

**Methods**

**Materials and chemicals.** Methylamine (MA) solution in ethanol, formamidine acetate, hydrobromic acid (HBr), hydroiodic acid (HI), lead bromide (PbBr$_2$), lead iodide (PbI$_2$), octylamine, anhydrous N,N-dimethylformamide (DMF), γ-butyrolactone, tetrahydrofuran (THF), chlorobenzene (CB) were all purchased from Sigma-Aldrich. PEDOT:PSS (AI 4083) was purchased from Heraeus. 1,3,5-tris(N-phenylbenzimiazole-2-yl)benzene (TPBi), cesium carbonate (Cs$_2$CO$_3$), 4,6-Bis(3,5-di(pyridin-3-yl)phenyl)-2-methylpyrimidine (B3PYMPM), Poly(4-butylphenyldiphenylamine) (poly-TPD) and Bis(1-phenylisoquinoline)(acetylacetonate)iridium(III) (Ir(piq)$_2$(acac)) were purchased from Lumtec. The colloidal zinc oxide (ZnO) nanoparticles (N-11) was purchased from Avantama. All chemicals were used as received.

**Lead-halide perovskite preparation.**

**MABr**: CH$_3$NH$_2$ solution in ethanol was reacted with HBr in water with excess methylamine at 0 °C. The precipitate was recovered via evaporation at 60 °C. The yellowish raw product was washed with diethyl ether for 5 times and dried at 60 °C in a vacuum oven for 24 h.

**MAPbBr$_3$ PeNCs:** 0.225 mmol MABr and 0.15 mmol PbBr$_2$ were dissolved in 0.5 mL anhydrous DMF and 8.5 mL γ-butyrolactone mixed solution with 15 µL octylamine to form a halide perovskite precursor solution. 0.3 mL of the perovskite precursor solution was dropped into 5 mL toluene with vigorous stirring. The colloidal solution was stored in sealed vial for one week before using. After ageing one week, we can get the perovskite NCs with higher PLQY. Then, the colloidal solution was centrifuged at 5000 rpm for 5 min to discard the aggregated precipitates and the PeNCs was obtained after centrifuged at 14,800 rpm. The PeNCs product was dispersed in solvent to form the colloidal solution with 5 mg/mL concentration for LEDs fabrication.

**MAPbBr$_3$ bulk crystal:** 1 mmol MABr and 1 mmol PbBr$_2$ were dissolved in 0.5 mL DMF and 0.5 mL γ-butyrolactone mixed solution and stored in open vial. The precursor solution was kept in oven at 60 °C for 24 h for crystal growing. The separated bulk crystals were collected for measurement (Supplementary Fig. 11).

**PeNCs characterization.** The structure of as-grown samples was characterized using X-ray diffraction (XRD, Bruker D8 Advanced Diffractometer with Cu Kα radiation) and transmission electron microscopy (TEM) with energy-dispersive X-ray spectroscopy (TEM; subångström-resolution, aberration-corrected TITAN G2 60-300, operated under 60 kV). UV-vis absorption was measured using LAMBDA 950 UV/Vis/NIR spectrophotometer and PL spectra were recorded through a micro-Raman spectrometer (Horiba, LABRAM HR800) with a single-grating setup and a 442 nm solid state laser were used as the excitation laser. The morphology and structure of the LED device were characterized by atomic force microscopy (AFM, Cypher ES SPM) in the contacting mode and field-emission scanning electron microscopy (FE-SEM, JEOL JSM-7001F). UPS measurements were performed on the samples in a home-built UHV multi-chamber system with base pressure better than $1 \times 10^{-9}$ torr. NMR was measured on Bruker 500MHz spectrometer and DMSO-D6 was used as solvent to dissolve the sample. The UPS source is from a helium discharge lamp (hv = 21.2 eV). The photoelectrons were measured by an electron analyzer (Omicron EA125).

**PLQY measurement.** PLQY measurements of colloidal PeNCs samples dispersed in THF solution and thin-film samples deposited onto sapphire substrate using the identical colloidal solution were carried out by coupling a Quanta-Phi integrating sphere to the Horiba Fluorolog system with optical fiber bundles. Both excitation and emission spectra were collected for the two cases of the sample directly illuminated by the excitation beam path in the integrating sphere and the empty sphere itself. A monochromatized Xe lamp was used as excitation source with wavelength of 400 nm and power of 1 mW cm$^{-2}$.

**Transient photoluminescence.** For femtosecond optical spectroscopy, the laser source was a Coherent Libra$^{TM}$ regenerative amplifier (50 fs, 1 KHz, 800 nm) seeded by a Coherent Vitesse$^{TM}$ oscillator (50 fs, 80 MHz). 800 nm wavelength laser pulses were from the regenerative amplifier while laser pulses at 400 nm were obtained with a BBO doubling crystal. The laser pulses (circular spot, diameter 3.5 mm) were directed onto the films. The emission from the samples was collected at a backscattering angle of 150° by a pair of lenses into an optical fiber that was coupled to a spectrometer (Acton, Spectra Pro 2500i) and detected by a charge coupled device (Princeton Instruments, Pixis 400B). The PL at different excitation intensity levels was recorded under the same exact experimental conditions except for the excitation intensity, which was controlled by adjusting the neutral density filters. Under the assumption of negligible nonlinear absorption and negligible sample damage, the relative PLQY at different excitation intensity was obtained by normalizing the collected PL intensity with the corresponding excitation light intensity.

**Transistor testing.** The transistors performance was measured to extract mobility with a Keithley 4200 semiconductor parameter analyzer in air at room temperature. A Si/SiO$_2$ substrate with patterned gold electrodes was selected as a substrate to make a bottom gate field effect transistor, and a PeNCs film was deposited on top as the active layer by spin-casting.

**PeLEDs fabrication.** The patterned ITO-glass substrates were cleaned with de-ionized water, acetone and isopropanol in sequence, followed by oven-drying at 80 °C for 2 h and O$_2$ plasma treatment (FEMTO) for 15 min. PEDOT:PSS was deposited onto the substrate by spin-casting at 4,000 rpm for 60 s, and annealed at 120 °C for 30 min in air. Colloidal PeNCs solution was spin-coated onto PEDOT:PSS to form the PeNCs layer. The remaining functional layers and aluminum electrodes were fabricated by thermal evaporation in a single run at a base pressure of <2 × 10$^{-4}$ Pa without breaking the vacuum. All the organics were evaporated at a rate of around 0.1-0.2 nm s$^{-1}$, and the aluminum electrodes were evaporated at a rate of 0.8-1 nms$^{-1}$. A shadow mask was used to define the cathode. All devices were completed with rudimentary encapsulation in glovebox for characterization in air at room temperature. The PEDOT:PSS film here serves as the ITO modification layer to enhance the injection and transport of hole. The pure TPBi and B3PYMPM film serve as electron transport layer (ETL). The TPBi:Cs$_2$CO$_3$ and B3PYMPM:Cs$_2$CO$_3$ film serve as the n-type doped electron injection layer (EIL) which can enhance the electron injection and decrease the driving voltage effectively[20]. All blended function layers were deposited using multi-source co-evaporation. The lowest unoccupied molecular orbital (LUMO) and the highest occupied molecular orbital (HOMO) of all materials were obtained from references[20,34,37]. The valance band and conductive band of PeNCs were obtained by ultraviolet photoelectron spectroscopy (UPS) and absorption spectrum (Supplementary Fig. 1, 3).

**Electron-only devices fabrication.** The devices structures are as follows: Device 1: ITO/ZnO (40 nm)/TPBi (30 nm)/TPBi:Cs$_2$CO$_3$ (10:1 w/w, 20 nm)/Al (100 nm). Device 2: ITO/ZnO (40 nm)/TPBi:B3PYMPM (2:1 w/w, 30 nm)/B3PYMPM:Cs$_2$CO$_3$ (10:1 w/w, 20 nm)/Al (100 nm). Device 3: ITO/ZnO (40 nm)/TPBi:B3PYMPM (1:2 w/w, 30 nm)/B3PYMPM:Cs$_2$CO$_3$ (10:1 w/w, 20 nm)/Al (100 nm). Device 4: ITO/ZnO (40 nm)/ B3PYMPM (30 nm)/B3PYMPM:Cs$_2$CO$_3$ (10:1 w/w, 20 nm)/Al (100 nm). Device with PeNCs layer: ITO/ZnO (40 nm)/ MAPbBr$_3$ PeNCs (40 nm)/B3PYMPM (30 nm)/B3PYMPM:Cs$_2$CO3 (10:1 w/w, 20 nm)/Al (100 nm). ZnO nanoparticle film was deposited onto the substrate by spin-casting at 3,000 rpm for 60 s, and annealed at 150 °C for 10 min in air. Except the deposition of ZnO nanoparticles film, the device fabrications were the same with PeLEDs.

**MAPbBr$_3$ PeNCs layer deposition.**

**Control devices:** A THF solvent dispersed PeNCs colloidal solution was selected for the PeNCs layer deposition with an optimized rotational speed of 1,500 rpm (Supplementary Table 2 and Supplementary Fig. 6).

**Champion devices:** A mixed solvent of THF:CB (16:1 v/v) was used to disperse PeNCs to obtain a uniform colloidal solution. A programmable spin-casting was developed for deposition of PeNCs film, which consists of two steps: step 1 was chosen as 400 rpm for 3 s; and step 2 was 1,500 rpm for 60 s.

**PeLEDs characterization.** With a nearly ideal Lambertian emission, the voltage independent EL spectra of the PeLEDs were measured using a Photo Research Spectra Scan spectrometer PR705 at different brightness, while the luminance-current-voltage characteristics of the devices were measured simultaneously with a programmable Agilent B2902A sourcemeter and a Konica-Minolta LS-110 luminance meter in air at room temperature (Supplementary Fig. 5b). The lifetime test was carried out by a programmable sourcemeter of Agilent B2902A and a calibrated Si-photodetector (Thorlabs, FDS-

1010CAL).

**PL of electron single carrier device:** PL spectra were recorded through a micro-Raman spectrometer (Horiba, LABRAM HR800) with a single-grating setup and a 442 nm solid state laser were used as the excitation laser. The bias was applied by Agilent B1500A semiconductor device parameter analyzer.

## Acknowledgements

This research is supported by the National Research Foundation, Prime Minister's Office, Singapore under its competitive Research Programme (CRP Award No. NRF-CRP14-2014-03). H.V.D. gratefully acknowledges the financial support from NRF Investigatorship grant NRF-NRFI2016-08 and additional support from TUBA. Q.H.X. acknowledges financial support from Singapore National Foundation via an Investigatorship Award (NRF-NRFI2015-03), and Singapore Ministry of Education through grants (MOE2015-T2-1-047 and MOE2015-T1-001-175). G.H.X. acknowledges financial support from the Science and Technology Development Fund from Macau SAR (FDCT-116/2016/A3) and Start-up Research Grant (SRG2016-00087-FST) from Research & Development Office at University of Macau.

## Author contributions

H.V.D., Q.H.X., and E.H.S. planned and guided the work. F.Y. designed and fabricated the devices, performed the device experiments, analyzed the data with assistance of S.T.T., and discussed the results with H.V.D. J.X designed and synthesized the perovskite materials and performed the material experiments and data analyses with assistance of J.X.Z, L.L.Z. and L.N.Q., and discussed the results with Q.H.X. The luminescence lifetime measurement made by R. S. The luminescence dynamics study was performed by G.C.X. The TEM measurement was made by Y.W.Z. The UPS measurement was made by S.C. and A. H. All authors discussed and commented on the manuscript.

## Additional information

## Competing financial interests

The authors declare no competing financial interests.

# Figure Caption

**Figure 1. Luminescence study and synthesis of CH₃NH₃PbBr₃ emitters. a,** The calculated injected carrier density dependence of luminescence quantum yield for excitonic emission ($k_1^{nr} = 10^7\ s^{-1}$, $k_1^{r} = 10^8\ s^{-1}$, $k_3^{nr} = 10^{-26}\ cm^6\ s^{-1}$) and bimolecular emission ($k_1^{nr} = 10^7\ s^{-1}$, $k_2^{r} = 10^{-10}\ cm^3\ s^{-1}$, $k_3^{nr} = 10^{-26}\ cm^6\ s^{-1}$) in different dimensional CH₃NH₃PbBr₃ emitters. **b,** Photon-injected charge carrier density dependence of the initial time PL intensity ($I_{PL}(t = 0)$). The quadratic behavior $I_{PL}[t=0]$ confirms the bimolecular recombination in bulk perovskite. **c,** Relative PLQY as a function of the photon-injected charge carrier density. **d,** PL spectra of colloidal PeNCs solution aging for 6 days. Peak emission wavelengths redshift: 506.5, 512.5, 514.6, 516.4, 517.8, 518.5 nm. The FWHM decreases: 32.7, 25.7, 24.9, 24.8, 24.7, 24.7 nm. **e,** XRD pattern migration of PeNCs in 6 days. **f,** TEM and high-resolution TEM image of the as-synthesized PeNCs. Scale bars: 50 and 2 nm. d (200) = 2.93 Å, d (210) = 2.65 Å.

**Figure 2. Device structure and L-I-V curves of devices. a,** Device structure with energy level alignment. **b,** SEM of device cross-section showing the multi-layer structure with distinct contrast. Scale bar: 100 nm. **c,** I-V curves of electron single carrier devices with different electron transport layers. **d,** Normalized PL spectrum of PeNCs film and normalized EL spectra of champion PeLEDs at different brightness levels from 500 to 15,000 cd m⁻² around. The peak emission wavelength is 524 nm and the FWHM is 22 nm. **e,** I-V curves of all PeLEDs. **f,** Luminance-Voltage dependences of devices.

**Figure 3. Comparison of devices. a,** EQE-current density characteristics. **b,** EQE-voltage dependences. **c,** Power efficiency-luminance correlations. **d,** The EQE$_{max}$ histogram for 15 champion devices from 4 batches. **e,** The dependencies between EQE$_{max}$ and the corresponding driving current density (scatters: experiment result; solid line: simulation by linear fitting). **f,** The PL intensity of MAPbBr3 layer in an electron single carrier device. $n_o$ is the density of electron injected into PeNCs layer when the current density is $1.0 \times 10^{-2}$ mA cm$^{-2}$ (scatters: experiment result; solid line: simulation by linear fitting).

**Table 1.** Key parameter comparison of PeLEDs.

| Device | Structure | $V_{on}$ (V) | $L_{max}$ (cd m$^{-2}$) | $EQE_{max}$ (%) | $\eta_{max}$ (lm W$^{-1}$) |
|---|---|---|---|---|---|
| Control device I | ITO / PEDOT:PSS (40 nm) / MAPbBr$_3$ PeNCs (40 nm) / TPBi (30 nm) / TPBi: Cs$_2$CO$_3$ (10:1 w/w, 20 nm) / Al (100 nm). | 3.3 | 15,130 | 5.11 | 10.82 |
| | | 3.4 | 16,470 | 5.06 | 7.03 |
| | | 3.6 | 14,190 | 4.81 | 7.49 |
| Control device II | ITO / PEDOT:PSS (40 nm) / MAPbBr$_3$ PeNCs (40 nm) / B3PYMPM (30 nm) / B3PYMPM: Cs$_2$CO$_3$ (10:1 w/w, 20 nm) / Al (100 nm). | 2.7 | 24,410 | 2.39 | 5.96 |
| | | 2.6 | 23,290 | 2.28 | 6.21 |
| | | 2.7 | 23,120 | 2.19 | 5.62 |
| Control device III | ITO / PEDOT:PSS (40 nm) / MAPbBr$_3$ PeNCs (40 nm) / B3PYMPM:TPBi (1:2 w/w, 30 nm) / B3PYMPM: Cs$_2$CO$_3$ (10:1 w/w, 20 nm) / Al (100 nm). | 3.0 | 33,570 | 7.91 | 20.18 |
| | | 2.9 | 30,380 | 6.75 | 17.92 |
| | | 3.0 | 32,730 | 7.25 | 18.47 |
| Champion device | ITO / PEDOT:PSS (40 nm)/ MAPbBr$_3$ PeNCs (80 nm) / B3PYMPM:TPBi (1:2 w/w, 30 nm) / B3PYMPM: Cs$_2$CO$_3$ (10:1 w/w, 20 nm) / Al (100 nm). | 3.1 | 22,830 | 12.9 | 26.07 |
| | | 3.0 | 43,400 | 7.53 | 17.56 |
| | | 2.9 | 25,410 | 10.67 | 30.3 |

$V_{on}$: turn on voltage @ 1 cd m$^{-2}$;

$L_{max}$: maximum luminance;

$EQE_{max}$: maximum external quantum efficiency;

$\eta_{max}$: maximum power efficiency.

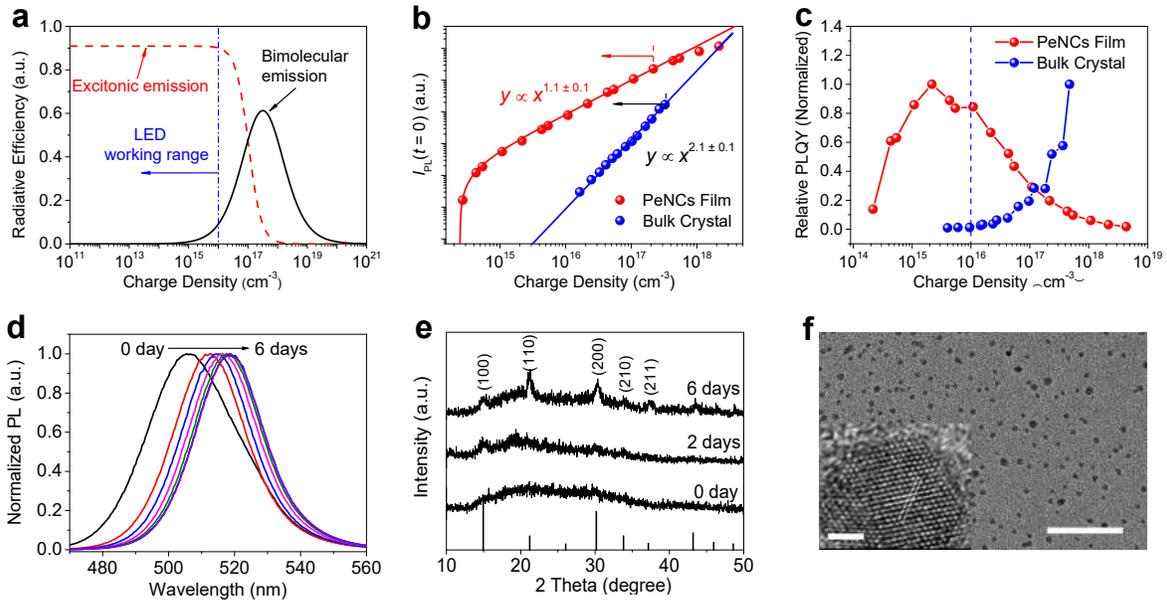

**Figure 1**

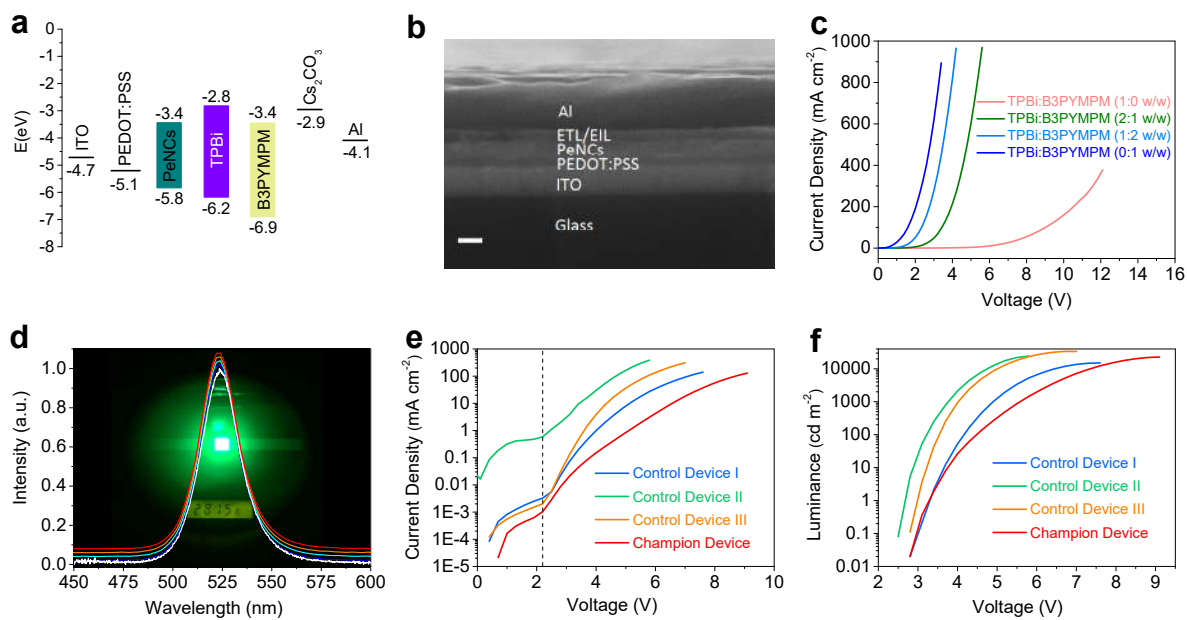

**Figure 2**
24

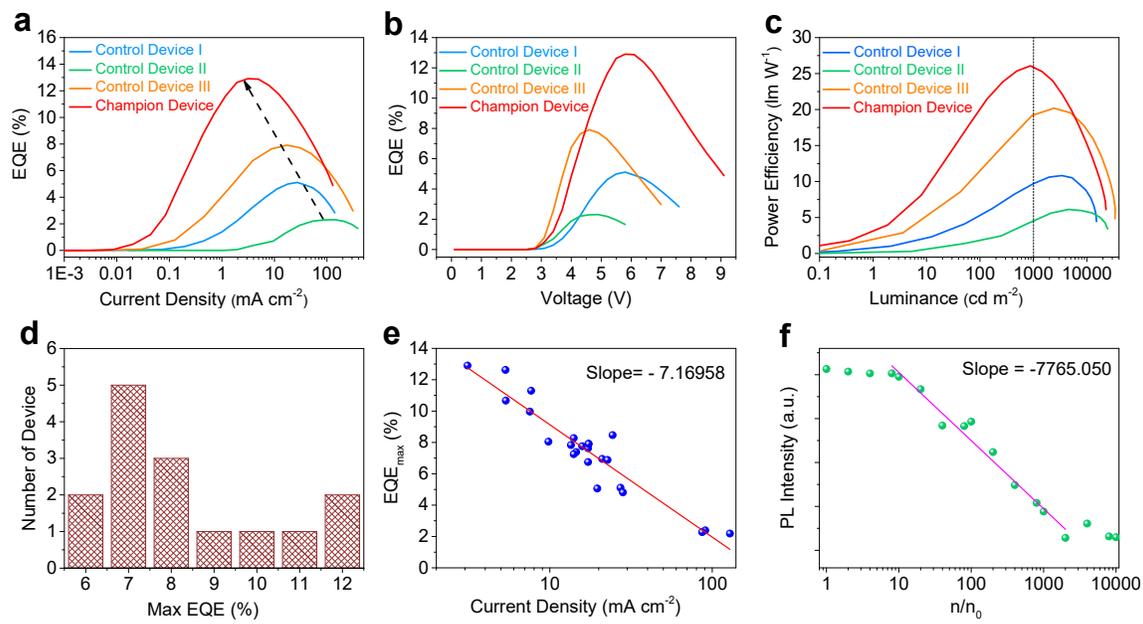

**Figure 3**